\documentclass[12pt]{article}

\usepackage{epsfig}

\usepackage{graphicx}

\def\square{\kern1pt\vbox{\hrule height 1.2pt
\hbox{\vrule width 1.2pt\hskip 3pt
\vbox{\vskip 6pt}\hskip 3pt\vrule width 0.6pt}
\hrule height 0.6pt}\kern1pt}
\def\ltwid{\mathrel{\raise.3ex\hbox{$<$\kern-.75em\lower1ex\hbox{$\sim$}}}}
\def\gtwid{\mathrel{\raise.3ex\hbox{$>$\kern-.75em\lower1ex\hbox{$\sim$}}}}

\begin{document}

\begin{titlepage}

\begin{flushright}
CCTP-2010-10 \\ UFIFT-QG-10-05
\end{flushright}

\vspace{0.5cm}

\begin{center}
\bf{Possible Enhancement of High Frequency Gravitational Waves}
\end{center}

\vspace{0.3cm}

\begin{center}
Maria G. Romania$^{\dagger}$ and  N. C. Tsamis$^{\ddagger}$
\end{center}
\begin{center}
\it{Department of Physics, University of Crete \\
GR-710 03 Heraklion, HELLAS.}
\end{center}

\vspace{0.2cm}

\begin{center}
R. P. Woodard$^{\ast}$
\end{center}
\begin{center}
\it{Department of Physics, University of Florida \\
Gainesville, FL 32611, UNITED STATES.}
\end{center}

\vspace{0.3cm}

\begin{center}
ABSTRACT
\end{center}
\hspace{0.3cm} We study the tensor perturbations 
in a class of non-local, purely gravitational models 
which naturally end inflation in a distinctive phase 
of oscillations with slight and short violations of 
the weak energy condition. We find the usual generic
form for the tensor power spectrum. The presence of
the oscillatory phase leads to an enhancement of 
gravitational waves with frequencies somewhat less 
than $10^{10} H\!z$.

\vspace{0.3cm}

\begin{flushleft}
\end{flushleft}

\vspace{0.1cm}

\begin{flushleft}
$^{\dagger}$ e-mail: romania@physics.uoc.gr \\
$^{\ddagger}$ e-mail: tsamis@physics.uoc.gr \\
$^{\ast}$ e-mail: woodard@phys.ufl.edu
\end{flushleft}

\end{titlepage}

${\bullet \; \;}$ {\bf Introduction:} 
During the inflationary era infrared virtual gravitons are 
ripped out of the vacuum by the accelerated expansion of 
spacetime \cite{Grishchuk}. We have proposed a mechanism 
through which the back-reaction to this process can slow 
the inflationary expansion rate \cite{NctRpw-1}. The idea 
is that the energy density of the newly produced gravitons 
comes not just from their bare kinetic energy but also from
the interactions they have with other gravitons. Because
gravity is attractive, these interactions act to slow 
inflation. If one imagines the process occurring on a closed 
spatial manifold such as $T^3$ it is easy to show that the 
energy density of this gravitational self-interaction grows 
as the exponential of twice the number of e-foldings, assuming 
{\it all} the gravitons on a 3-surface were in interaction 
with one another \cite{NctRpw0}. That would be make the 
manifold suffer gravitational collapse after only about 14 
e-foldings! Of course this estimate ignores causality; any 
given graviton only feels the potentials from the gravitons 
which are visible within its past light-cone. The actual 
growth of the interaction energy is therefore like the number 
of e-foldings, so the system requires something like $10^6$ 
e-foldings of inflation to evolve itself to the point of 
gravitational collapse. However, it seems inevitable that 
this effect must eventually stop inflation if nothing else 
supervenes.
 
The mechanism just sketched provides a wonderful way of 
resolving two perplexing problems:
\begin{itemize}
\item{Explaining why the dimensionless product of Newton's 
constant $G$ and the cosmological constant $\Lambda$ is 
about $10^{-122}$; and}
\item{Finding a natural model of inflation.}
\end{itemize}
Our solution to the first problem is to deny its premise: 
we do not believe $\Lambda$ is actually small. It only 
seems to be small when one infers its value from the current 
expansion rate using classical gravity, which ignores the 
quantum gravitational screening mechanism. To be specific, 
we believe $G \Lambda \ltwid 10^{-6}$ is just small enough 
to justify using the semi-classical approximation, and that 
this is what started primordial inflation, without the need 
for a scalar inflaton. This gives a very attractive model 
of inflation because:
\begin{itemize}
\item{It dispenses with the need to assume that the initial 
configuration of the inflaton is homogeneous over more than 
a Hubble volume since the cosmological {\it constant} is 
homogeneous over {\it all} space;}
\item{The long period of inflation required for back-reaction 
to build up dispenses with the problem of having to fine tune 
the inflaton potential to make inflation last long enough;}
\item{The screening mechanism dispenses with the need to fine 
tune the inflaton potential so that inflation ends with nearly 
zero cosmological constant;}
\item{The fact that inflation is driven by the bare cosmological
constant, rather than a scalar inflaton, dispenses with the need 
to invent a new, otherwise undetected field; and}
\item{The fact that this model contains only the single 
dimensionless parameter $G \Lambda$ means that it makes 
unique predictions, unlike scalar-driven inflation which 
can be tuned to give almost any result for the scalar power 
spectrum.}
\end{itemize}

However, there is a severe problem of {\it tractability} when 
it comes to using this model. The bare kinetic energy of 
inflationary gravitons is a one-loop effect, so the interaction 
energy we seek does not appear until two-loop order. Two-loop 
results are not easy to compute even for simple theories on 
flat space background; this one requires quantum gravity on 
de Sitter background. In fact, the graviton 1-point function 
has been evaluated at two-loop order and it does seem to support 
the relaxation mechanism \cite{NctRpw1}, but this computation 
required a year's labor and its interpretation is open to debate 
owing to the difficulty of formulating an invariant definition 
of relaxation \cite{inv}. Furthermore, at the same time as
the two-loop effect finally becomes significant, the effects 
of higher loops also become significant. So one requires a 
non-perturbative resummation technique to evolve into the late 
time regime at which interesting predictions can be made.

We believe it may be possible to derive such a non-perturbative 
resummation technique by extending the stochastic method which
Starobinsky devised for the same purpose in scalar potential 
models \cite{AAS,NctRpw1.5}. However, generalizing this technique 
to gravity is a difficult problem \cite{Miao}. This paper is part 
of a parallel effort \cite{NctRpw2} which is based on the idea 
of {\it guessing} the most cosmologically significant part of 
the effective field equations of quantum gravity. While there 
is no chance of guessing the full effective field equations, 
it might be possible to guess just enough to correctly describe 
the evolution of the scale factor $a(t)$ for a homogeneous and 
isotropic geometry, using what we know from perturbation theory 
about how the back-reaction effect scales.

An important feature of any model which relaxes the cosmological 
constant is non-locality. Nonlocal models of cosmology have been 
much studied \cite{NctRpw0,nonlocal} because they can avoid the 
problem that de Sitter must be a solution for any local, stable 
theory, and because non-local couplings between different times 
can ease fine tuning problems. In a previous paper \cite{NctRpw2} 
we proposed a phenomenological model which can provide evolution 
beyond perturbation theory. In one sentence, we constructed an 
{\it effective} conserved stress-energy tensor $T_{\mu\nu} [g]$ 
which modifies the gravitational equations of motion:
\footnote{Hellenic indices take on spacetime values while 
Latin indices take on space values. Our metric tensor 
$g_{\mu\nu}$ has spacelike signature and our curvature 
tensor equals:
$R^{\alpha}_{~\beta\mu\nu} \equiv 
\Gamma^{\alpha}_{~\nu\beta, \mu} +
\Gamma^{\alpha}_{~\mu\rho} \;
\Gamma^{\rho}_{~\nu\beta} -
(\mu \leftrightarrow \nu)$. 
The initial Hubble parameter is $3H^2_0 \equiv \Lambda$. 
We restrict our analysis to scales 
$M \equiv (\, \Lambda / 8 \pi G \,)^{\frac14}$ 
below the Planck mass $M_{\rm Pl} \equiv G^{-\frac12}$ 
so that the dimensionless coupling constant $G \Lambda =
(\, M / M_{\rm Pl} \,)^4 \,$ of the theory is small.}
\begin{equation}
G_{\mu\nu} \; \equiv \;
R_{\mu\nu} \, - \, \frac12 g_{\mu\nu} \, R \; = \;
- \Lambda \, g_{\mu\nu} \, + \,
8 \pi G \, T_{\mu\nu}[g] 
\;\; . \label{eom1}
\end{equation}
and which, we hope, contains the most cosmologically
significant part of the full effective quantum
gravitational equations.

What form to guess for $T_{\mu\nu}[g]$ was motivated by 
what we seek to do, and by what we know from perturbation 
theory. We seek to describe cosmology, which implies 
homogeneous and isotropic geometries. When specialized to 
such a geometry the full effective stress tensor must take 
the perfect fluid form and we lose nothing by assuming 
that generally:
\begin{equation}
T_{\mu\nu}[g] \; = \;
(\rho + p) \, u_{\mu} \, u_{\nu} \, + \,
p \, g_{\mu\nu}
\;\; . \label{Tmn}
\end{equation}
The relation between $p[g]$, $\rho[g]$ and $u_{\mu}[g]$ is 
heavily constrained by stress-energy conservation, but it is
possible to specify one function for free. It turns out to be
computationally simplest to take this free function to be the
pressure \cite{NctRpw2}. We further require the pressure to be 
an ordinary function of some non-local scalar which grows like 
the number of e-foldings when specialized to de Sitter. The 
simplest choice for such a scalar is the inverse d'Alembertian 
acting on the Ricci scalar \cite{NctRpw0}. If the pressure is 
to grow the way we know it does from perturbation theory 
\cite{pert}, and to eventually end inflation, then one is 
lead to the form \cite{NctRpw2}:
\begin{equation}
p[g](x) \; = \;
\Lambda^2 \, f[- G \Lambda \, X](x) 
\qquad , \qquad
X \, \equiv \, \frac{1}{\square} \, R
\;\; , \label{pressure}
\end{equation}
where the function $f$ grows without bound and satisfies:
\begin{equation}
f[- G \Lambda \, X] \; = \;
- G \Lambda \, X \, + \, 
O\Big[ (G \Lambda)^2 \Big]
\;\; , \label{fincr}
\end{equation}
and where the scalar d'Alembertian:
\begin{equation}
\square \, \equiv \;
\frac{1}{\sqrt{-g}} \;
\partial_{\mu} \Big( \,
g^{\mu\nu} \sqrt{-g} \; \partial_{\nu} \, \Big)
\;\; , \label{box}
\end{equation}
is defined with retarded boundary conditions.
The induced energy density $\rho[g]$ and 4-velocity
$u_{\mu}[g]$ are determined, up to their initial
value data, from stress-energy conservation:
\begin{equation}
D^{\mu} \, T_{\mu\nu} \; = \; 0
\;\; . \label{cons1}
\end{equation}
The 4-velocity was chosen to be timelike and
normalized:
\begin{equation}
g^{\mu\nu} \, u_{\mu} u_{\nu} = -1
\qquad \Longrightarrow \qquad
u^{\mu} \, u_{\mu ; \nu} = 0
\;\; . \label{u}
\end{equation}

The homogeneous and isotropic evolution
\footnote{The line element in co-moving coordinates is 
$ds^2 = -dt^2 + a^2(t) \, d{\vec x} \cdot d{\vec x}$.
In terms of the scale factor $a$, the Hubble parameter 
equals $H(t) = {\dot a} \, a^{-1}$ and the deceleration 
parameter equals $q(t) = - a \, {\ddot a} \, {\dot a}^{-2} 
= -1 - {\dot H} \, H^{-2} \equiv -1 + \epsilon(t)$.}
of this model -- using a combination of numerical 
and analytical methods -- revealed the following 
basic features: 
\footnote{In \cite{NctRpw2}, our analytical results 
were obtained for any function $f$ satisfying 
(\ref{fincr}) and growing without bound, our numerical 
results for the choice: $f(x) = \exp(x) - 1$.} 
\\
-- After the onset and during the era of inflation, 
the source $X(t)$ grows while the curvature scalar 
$R(t)$ and Hubble parameter $H(t)$ decrease. \\ [5pt]
-- Inflationary evolution dominates roughly until 
we reach a critical point $X_{cr}$ defined by:
\begin{equation}
1 - 8 \pi G \Lambda \, f[ - G \Lambda \, X_{cr}] 
\; \equiv \; 0
\;\; . \label{Xcr}
\end{equation}
-- The epoch of inflation ends close to but before 
the universe evolves to the critical time. This is 
most directly seen from the deceleration parameter 
since initially $q(t=0) = -1$ while at criticality 
$q(t=t_{cr}) = +\frac12$. \\ [5pt]
-- Oscillations in $R(t)$ become significant as we 
approach the end of inflation; they are centered 
around $R = 0$, their frequency equals:
\begin{equation}
\omega \; = \; 
G \Lambda H_0 \sqrt{72 \pi \, f_{cr}'}
\;\; , \label{omega}
\end{equation}
and their envelope is linearly falling with time. 
\\ [5pt]
-- During the oscillations era, although there is 
net expansion, the oscillations of $H(t)$ take it
to small negative values for short time intervals --
a feature conducive to rapid reheating; those of 
${\dot H}(t)$ take it to positive values for about 
half the time; and, those of $a(t)$ are centered 
around a linear increase with time.

Furthermore, this simple model can be improved to 
account for the requirement to naturally produce 
negative pressure of the correct magnitude during 
the current epoch \cite{NctRpw3}. However, for the 
purposes of this paper such an improvement is
inconsequential since we shall be concerned with
the period about the end of inflation where the
simple model and its improvement are indistinguishable.

A novel feature of this class of models is the 
existence of an oscillatory regime of short duration
which commences towards the very end of the inflationary 
era.  During this period ${\dot H}(t)$ is positive
about half the time, which represents a violation of 
the weak energy condition. Such a violation cannot 
occur in classically stable theories \cite{Cline} but 
it can be driven by quantum effects of the type we
seek to model without endangering stability. 

A simple, fully worked-out example is provided by a 
massless, minimally coupled scalar with a $\phi^4$ 
potential on non-dynamical de Sitter background. For 
that model, inflationary particle production tends to 
push the scalar up its potential, which of course increases
the vacuum energy and leads to a violation of the weak
energy condition. That has been confirmed in a fully
renormalized computation of the expectation value of
the model's stress tensor at two-loop order \cite{phi4I}. 
That there is no instability was confirmed by a fully
renormalized computation of the scalar self-mass-squared
at two-loop order, which was then used to solve the
linearized effective field equations \cite{phi4II}. As 
might be expected, pushing the scalar up its potential
makes it develop a small mass, which actually makes the 
system more stable, not less. And a fully non-perturbative
analysis by Starobinsky and Yokoyama confirms that the
system approaches a static limit \cite{phi4III}. 

Our own model is an attempt to model the most cosmologically
significant features of the inflationary production of
gravitons, so it shares many features with the simpler 
$\phi^4$ model. Of course it is quite a bit more difficult 
derive comparably powerful results for quantum gravity.
However, we have been able to show that there are no tachyonic
modes \cite{NctRpw4}.

The oscillatory phase and the associated weak energy condition
violation is a very distinctive feature of our model and 
cannot occur in classical, scalar-driven inflation. The 
purpose of this study is to determine whether this oscillatory 
regime leaves its signature on the observable tensor power 
spectrum.
\footnote{The more complicated analysis of the scalar 
power spectrum that this class of non-local cosmological
models predicts has been done elsewhere \cite{NctRpw4}.}
We shall, therefore, obtain the amplitude and frequency 
of two kinds of gravitational waves and examine their 
evolution under the expansion history that this class 
of models predicts. The first kind of waves is now of 
cosmological scale and originated during inflation while 
the second kind was on the verge of experiencing first 
horizon crossing when the epoch of oscillations began. 
Because we shall be relating scales from the very early 
universe to current measurements, we first focus on 
presenting the basic equations and relevant relations, 
and then we apply them for our purpose. \\

${\bullet \; \;}$ {\bf The Set-up:} 
The analysis of tensor perturbations in this class of 
models is much simpler than that of scalar perturbations 
\cite{NctRpw4}. The reason is that -- unlike the case of 
scalar perturbations -- the non-local nature of the model 
does not alter the basic equation which tensor perturbations 
$h_{ij}^{TT}$ satisfy at linearized order:
\begin{equation}
\left[ \, \frac{\partial^2}{\partial t^2} \, + \,
3 H(t) \, \frac{\partial}{\partial t} \, - \,
\frac{\nabla^2}{a^2(t)} \, \right] \,
h_{ij}^{TT} (t, x) \; = \; 0
\;\; . \label{basiceqn}
\end{equation}
Therefore, up to sub-dominant corrections coming from
the exact form of the mode functions before and after
first horizon crossing, the resulting power spectrum 
${\Delta}^2_h$ will have the usual form:
\begin{equation}
{\Delta}^2_h (k) \; \simeq \;
\frac{16}{\pi} \, G H^2 (t_k)
\;\; , \label{Deltah}
\end{equation}
where the Hubble parameter $H$ is evaluated at the
time $t_k$ of first horizon crossing of the mode with
wavenumber $k$:
\begin{equation}
k \; = \; H(t_k) \, a(t_k)
\;\; . \label{HorCross}
\end{equation}
Moreover, the tensor spectral index $n_T$ is defined 
as:
\begin{equation}
n_T \; \equiv \;
\frac{d}{d \ln k} \ln[{\Delta}^2_h (k)]
\;\; , \label{nT}
\end{equation}
and for the power spectrum (\ref{Deltah}) equals:
\begin{equation}
n_T \; \simeq \;
- \frac{2 \epsilon(t_k)}{1 - \epsilon(t_k)}
\; \simeq \;
- 2 \epsilon(t_k)
\;\; , \label{nT2}
\end{equation}
where the last relation assumes that $\epsilon(t_k)
\ll 1$. It is apparent that knowledge of the relevant 
scale factor $a(t)$ suffices to compute the tensor 
power spectrum and spectral index. \\

${\bullet \; \;}$ {\bf Assumptions about the Expansion
History:} 
For the purposes of this paper we divide cosmological 
history into three epochs: \\ [-11pt]

{\it -- Primordial Inflation.} The most convenient time 
parameter for the epoch of primordial inflation is the 
number $N$ of e-foldings before criticality:
\begin{equation}
a(t) \; \equiv \; a_{\rm cr} \, e^{-N}
\;\; . \label{Ninfl}
\end{equation}
The important cosmological parameters during this 
phase are \cite{NctRpw4}:
\begin{eqnarray}
H^2(t) &\!\! \simeq \!\!& 
\frac19 \, \omega^2 \, \Big( 4N + \frac43 \, \Big) 
\;\; , \label{Hinfl} \\ 
\epsilon(t) &\!\! \simeq \!\!& 
\frac{2}{4 N + \frac43}
\;\; . \label{epsiloninfl}
\end{eqnarray}
At the end of inflation the scale factor is about 
$a_{\rm cr}$ and the Hubble parameter is about 
$\omega$. \\ [-11pt]

{\it -- Oscillations.} The distinctive feature of our 
model is the epoch of oscillations. The most convenient
time parameter during this era is the co-moving time 
after criticality:
\begin{equation}
\Delta t \; \equiv \;
t - t_{\rm cr}
\;\; . \label{Deltat}
\end{equation}
The important cosmological parameters during this phase 
are \cite{NctRpw2}:
\begin{eqnarray}
a(t) &\!\! \simeq \!\!&
a_{\rm cr} \, \Big[ \, \omega \, \Delta t \, + \,
1 \, + \, 
\sqrt2 \, \Big( \cos(\omega \, \Delta t) - 1 \Big) 
\, \Big]
\;\; , \label{aosc} \\
H(t) &\!\! \simeq \!\!&
\frac{\omega \, \Big[ \, 1 - 
\sqrt2 \, \sin(\omega \, \Delta t) \, \Big]}
{\omega \, \Delta t + ( 1 - \sqrt2 ) +
\sqrt2 \, \cos(\omega \, \Delta t)}
\;\; , \label{Hosc} \\
\epsilon(t) &\!\! \simeq \!\!&
\frac{\sqrt2 \, \Big[ \, 
\omega \, \Delta t + ( 1 - \sqrt2 ) \, \Big]
\cos(\omega \, \Delta t) \, + \, 3 \, - \,
2 \sqrt2 \, \sin(\omega \, \Delta t)}
{\Big[ \, 1 - \sqrt2 \, \sin(\omega \, \Delta t) \, \Big]^2}
\;\; . \qquad \label{epsilonosc}
\end{eqnarray}
The epoch of oscillations is terminated by the flow of 
energy density to the matter sector from the vast reservoir 
of super-horizon scalar modes, all of which begin to oscillate 
with frequency $\omega$. As explained in our analysis of scalar
perturbations \cite{NctRpw4}, we believe these oscillations 
should lead to very rapid reheating. 
Let us call the number of oscillatory e-foldings 
$\Delta N$. Then, at the end of the oscillations era:
\begin{equation}
a \; \simeq \; 
a_{\rm cr} \, e^{\Delta N}
\qquad , \qquad
H \; \simeq \;
\omega \, e^{-\Delta N}
\;\; . \label{oscend}
\end{equation}

Before proceeding, it is worth reviewing the argument -- in
section 6 of \cite{NctRpw4}-- for why the oscillatory phase 
should lead to rapid reheating. First, note that causality 
imposes absolutely no obstacle to the decay, eventually into 
radiation, of a super-horizon mode which is oscillating at 
a frequency comparable to the Hubble scale. Indeed, the usual 
mechanism of reheating relies on precisely such a decay of 
the inflaton zero mode, which has infinite frequency.
\footnote{See section 5.5 of the text by Mukhanov \cite{Slava}.} 
Second, note that the quantum gravitational back-reaction to 
inflation, which our non-local field equations seek to model, 
requires a very long period of nearly de Sitter inflation 
before enough inflationary gravitons have been created that 
their accumulated interaction is enough to slow inflation. 
One can estimate the number of inflationary e-foldings to 
be about $(G \Lambda)^{-1} \gtwid 10^6$. Third, this long 
period of inflation means that the number density of modes 
which have undergone first horizon crossing is {\it staggering} 
\cite{NctRpw4}:
\begin{equation}
n \, \sim \, 
\frac{H^3}{3 \pi^2} \, \exp\Bigl(\frac3{G \Lambda}\Bigr) 
\, \gtwid \, H^3 \times 10^{10^6} 
\;\; . \label{BIG}
\end{equation}
Finally, recall that in our model, {\it all} of these 
super-horizon modes begin oscillating at about the same 
time. In the face of a number density such as (\ref{BIG}) 
it makes little sense to attemp to estimate the rate at 
which the disturbance of a single oscillating mode would 
lead to the production of relativistic matter. As long as 
that rate is non-zero -- and the universal character of 
the gravitational coupling ensures it is non-zero -- then 
the vast number of modes which participate must make 
reheating almost instantaneous. \\ [-11pt]

{\it -- $\Lambda$CDM}. The $\Lambda$CDM cosmology after 
the epoch of oscillations is standard, and we do not 
require explicit forms for the three geometrical parameters. 
To compare quantities from the first two eras with their 
redshifted descendants at present time it is useful to 
express the energy density $\rho_R$ at the onset of the 
$\Lambda$CDM epoch in terms of the reheating temperature
$T_R$ and the number $n \approx 10^3$ of relativistic 
species:
\begin{equation}
\rho_R \; \simeq \;
\frac{3 c^2 \omega^2 \, e^{-2 \Delta N}}{8\pi G} 
\; \simeq \;
n \times 
\frac{\pi^2}{30} \frac{(k_B T_R)^4}{(\hbar c)^3} 
\;\; . \label{rhoR}
\end{equation}
The current energy density $\rho_{\rm now}$ can 
be written in terms of its tiny radiation fraction 
$\Omega_r \approx 8.5 \times 10^{-4}$ and the 
corresponding temperature $T_{\rm now} \approx 
2.726 K$ of that radiation \cite{WMAP}:
\begin{equation}
\rho_{\rm now} \; = \;
\frac{3 c^2 H_{\rm now}}{8\pi G} 
\; \simeq \;
\frac2{\Omega_{r}} \times 
\frac{\pi^2}{30} \frac{(k_B T_{\rm now})^4}{(\hbar c)^3} 
\;\; . \label{rhonow}
\end{equation}
Dividing (\ref{rhonow}) by (\ref{rhoR}) gives a relation 
between current conditions and those prevailing at the 
end of inflation:
\begin{equation}
\left( \frac{T_{\rm now}}{T_R} \right)^4 
\; \simeq \;
\frac{n \, \Omega_r}{2} \; e^{2 \Delta N}
\left( \frac{H_{\rm now}}{\omega} \right)^2
\;\; . \label{nowthen}
\end{equation}
We define $N_{\rm now}$ as the number of e-foldings 
from criticality to the present time. Using the relation 
(\ref{nowthen}) and:
\begin{equation}
\frac{T_{\rm now}}{T_R} 
\; \simeq \;
\frac{a_R}{a_{\rm now}} 
\; \simeq \;
\frac{a_{\rm cr} \, e^{\Delta N}}{a_{\rm now}}
\;\; , \label{Tnowthen}
\end{equation}
we see that $N_{\rm now}$ equals:
\begin{equation}
N_{\rm now} \; \simeq \;
\Delta N \, + \,
\ln \! \left[ \frac{T_R}{T_{\rm now}} \right] 
\; = \;
\frac12 \ln \! \left[ \frac{\omega}{H_{\rm now}} \right] 
\, + \, \frac12 \, \Delta N 
\, - \, \frac14 \ln \Bigl[ 2 n \, \Omega_r \Bigr]
\;\; . \label{Nnow}
\end{equation}
We shall later show that the measured value of the 
scalar power spectrum $\Delta^2_{\cal R}$ , and the 
current limit on the tensor-to-scalar ratio $r$, 
together imply the restriction $\, \omega \ltwid 
10^{55} H_{\rm now}$ . Hence we conclude:
\begin{equation}
N_{\rm now} \; \ltwid \;
63 \, + \, \frac12 \, \Delta N
\;\; . \label{Nnowval}
\end{equation}

The argument that $\omega \ltwid 10^{55} H_{\rm now}$ 
results from comparing expression (\ref{Deltah}) with
the current bound on the tensor contribution to the 
quadrupole moment. To make this comparison we must
solve the following problem concerning the 
relation between late times and early times: \\ [2pt]
{\it 01. Given a physical wave number $K_{\rm now}$ 
at the current time, find the e-folding $N_{\rm hor}$ 
when it experienced first horizon crossing during 
inflation.}
To solve this problem, we first use the horizon 
crossing condition (\ref{HorCross}) to express 
$K_{\rm now}$ in terms of $N_{\rm hor}$ :
\begin{eqnarray}
K_{\rm now} &\!\! = \!\!& 
\frac{k}{a_{\rm now}} 
\; = \;
\frac{k}{a(t_k)} \times \frac{a(t_k)}{a_{\rm cr}} 
\times \frac{a_{\rm cr}}{a_{\rm now}} 
\\
&\!\! \approx \!\!&
\frac13 \, \omega \, \sqrt{4 N + \frac43}
\, \times \, e^{-N_{\rm hor}} 
\, \times \, e^{-N_{\rm now}} 
\;\; . \label{K_now_then}
\end{eqnarray}
Now invert (\ref{K_now_then}) to solve for 
$N_{\rm hor}$ :
\footnote{The inversion was done under the assumption 
that the Hubble parameter at the end of inflation is 
much bigger than its present value: $\omega \,
H^{-1}_{\rm now} \gg 1$.}
\begin{eqnarray}
N_{\rm hor} &\!\! \approx \!\!& 
\ln \! \left[ \frac{\omega}{K_{\rm now}} \right] 
\, - \, N_{\rm now} \, + \, 
\frac12 \, \ln \! \left[ \, 
\frac49 N + \frac4{27} \, \right] 
\label{Nhor1} \\
&\!\! \approx \!\!& 
\frac12 \, \ln \! \left[ 
\frac{\omega H_{\rm now}}{c^2 K_{\rm now}^2} \right]
\, - \, \frac12 \, \Delta N \, + \,
\frac14 \, \ln \Bigl[ 2n \, \Omega_r \Bigr] \, + \,
\frac12 \, \left[ \, \frac29 \ln \Big( 
\frac{\omega H_{\rm now}}{c^2 K_{\rm now}^2} 
\Big) \right] 
\qquad \label{Nhor2}
\end{eqnarray}
For the $\ell$-th partial wave contribution to the 
anisotropies of the cosmic ray microwave background, 
the corresponding number $N_{\ell}$ of e-foldings 
before the end of inflation is:
\begin{eqnarray}
& \mbox{} &
\hspace{-1cm}
K_{\rm now} \; \approx \;
\frac{\ell}{2} \times \frac{H_{\rm now}}{c}
\quad \Longrightarrow 
\label{Nell} \\
& \mbox{} &
\hspace{-1cm}
N_{\ell} \; \approx \;
- \ln \Big( \frac{\ell}{2} \Big) + 
\frac12 \ln \Big( \frac{\omega}{H_{\rm now}} \Big) - 
\frac12 \Delta N + 
\frac14 \ln \Big( 2n \, \Omega_r \Big) + 
\frac12 \ln \! \left[ \, \frac29 \ln \Big(
\frac{\omega}{H_{\rm now}}
\Big) \right] 
\nonumber
\end{eqnarray}
The restriction $\, \omega \ltwid 10^{55} H_{\rm now}$ 
then implies:
\begin{equation}
N_{\ell} \; \ltwid \;
65 \, - \, 
\frac12 \ln \Big( \frac{\ell}{2} \Big) \, - \,
\frac12 \, \Delta N
\;\; . \label{Nellvalue}
\end{equation}
We cannot hope to detect a signal outside the range 
$2 \leq \ell \ltwid 100$, so the interesting values 
of $N_{\ell}$ lie within a band of only four e-foldings.

We now deduce the restriction on $\omega$ coming
from the measured value of the scalar power spectrum
$\Delta^2_{\cal R}$ \cite{WMAP}:
\begin{equation}
\Delta^2_{\cal R} (k_0) \; \approx \;
2.44 \times 10^{-9} 
\qquad , \qquad
k_0 \, \equiv \, 0.0002 \; (Mpc)^{-1}
\;\; , \label{DeltaRvalue}
\end{equation}
and the $95\%$ confidence bound on the tensor-to-scalar 
ratio: $r(k_0) \ltwid 0.22 \,$ \cite{WMAP}. Employing 
expressions (\ref{Deltah}) and (\ref{Hosc}) we get:
\begin{eqnarray}
\Delta^2_h (k_0) \; = \;
r(k_0) \; \Delta^2_{\cal R} (k_0)
&\!\! \simeq \!\!&
\frac{16}{9\pi} \,
G \omega^2 \Big( 4 N_0 + \frac43 \, \Big)
\label{Deltahvalue1} \\
&\!\! \ltwid \!\!&
\Big[ 0.22 \Big] \times 
\left[ 2.44 \times 10^{-9} \right]
\;\; . \label{Deltahvalue2}
\end{eqnarray}
Now the wave number $k_0$ and its associated number
of e-foldings $N_0$ correspond to the $\ell = 2$
partial wave so that -- under the asumption that
the Hubble parameter at the end of inflation is much
bigger than its present value ($\omega H_{\rm now}^{-1}
\gg 1$) and that the duration of the oscillations era 
is very short ($\Delta N < 10$) -- equation (\ref{Nell})
implies: $\, N_0 = N_{\ell = 2} \gtwid 60$. Thus,
expressions (\ref{Deltahvalue1}-\ref{Deltahvalue2})
reduce to $\, \omega \sqrt{G} \ltwid 2 \times 10^{-6}$ 
and when we convert to $H \! z$ we get:
\begin{equation}
\omega \; \ltwid \;
2 \times 10^{-6} \;
\sqrt{\frac{c^5}{G \hbar}}
\; \approx \;
3.7 \times 10^{37} H \! z
\quad \Longrightarrow \qquad
\omega \; \ltwid \;
10^{55} \, H_{\rm now}
\;\; , \label{omegabound}
\end{equation}
where we used $\, H_{\rm now} \approx 3.2 \times 
10^{-18} H \! z$ for the current value of the Hubble 
parameter. 

At this stage it is natural to consider a second 
problem which is in some ways the inverse 
of the first: \\ [2pt]
{\it 02. Given a physical wave number $K_N$ from the 
epoch of inflation, find its physical wave number now.} 
To achieve this, we express the current physical wave 
number in terms of $K_N$ and $N$:
\begin{eqnarray}
K_{\rm now} &\!\! = \!\!&
\frac{k}{a_{\rm now}} 
\; = \;
\frac{k}{a(t)} \times \frac{a(t)}{a_{\rm cr}} 
\times \frac{a_{\rm cr}}{a_{\rm now}} 
\; = \;
K_N \times e^{-N} \times e^{-N_{\rm now}}
\nonumber \\
&\!\! \approx \!\!&
\sqrt{\frac{K_N^2 \, H_{\rm now}}{\omega}} \, 
\left( \frac{n \, \Omega_r}{2} \right)^{\frac14} \,
e^{-N -\frac12 \Delta N} 
\;\; , \label{K_then_now}
\end{eqnarray}
where in the last step we used (\ref{Nnow}) for 
$N_{\rm now}$ . An important special case is the 
oscillation frequency $f_{\rm peak}$ for the wave
vector $\, K_N = \frac{\omega}{c} \,$ at $\, N = 
0 \,$. In this situation, (\ref{K_then_now}) gives:
\begin{equation}
f_{\rm peak} \; \equiv \; \frac1{2\pi} 
\left( \frac{c k}{a_{\rm now}} 
\right)_{K_0 = \frac{\omega}{c}} 
\; \approx \;
\frac{\sqrt{\omega \, H_{\rm now}}}{2 \pi} \,\, 
e^{-\frac12 \Delta N} 
\;\; . \label{freqnow}
\end{equation}
Imposing the restriction $\, \omega \ltwid 
10^{55} H_{\rm now}$ and using the current value of
$\, H_{\rm now} \approx 71 \, km \, s^{-1} (Mpc)^{-1} 
\approx 3.2 \times 10^{-18} H \! z$ \cite{WMAP}, 
implies:
\begin{equation}
f_{\rm peak} \; \ltwid \;
\Bigl( 10^{9} \, H \! z \Bigr) \, 
e^{-\frac12 \Delta N}
\;\; . \label{freqbound}
\end{equation}
Therefore, the late time descendant of the gravitational
waves produced by the phase of oscillations are unobservable
in the cosmic microwave background and have relevance only 
for high frequency direct detectors. \\

${\bullet \; \;}$ {\bf Overview of Gravitational 
Waves in the Oscillating Regime:} 
In terms of the mode functions $u(t, k)$ the basic 
equation (\ref{basiceqn}) takes the form:
\begin{equation}
{\ddot u} \, + \, 3 H \, {\dot u} \, + \,
\frac{k^2}{a^2} \, u \; = \; 0
\;\; . \label{ueqn}
\end{equation}
We do not possess exact forms for the two, linearly 
independent solutions during the oscillatory regime. 
Even if we had these solutions, we would not know the 
linear combination of them that gives ``the'' mode 
function $u(t,k)$, which we define to be the coefficient 
of the annihilation operator in the free field expansion 
of the graviton. It makes sense to first develop a 
reasonable approximation for the linearly independent 
solutions and then consider which combination of them 
occurs in the actual mode function $u(t,k)$. In 
approximating the solutions it also makes sense to 
first include the effect of the overall linear expansion 
-- for which exact solutions exist -- and then numerically
superimpose the effect of the oscillations. \\

${\bullet \; \;}$ {\bf The Case of Linear Expansion:}
During the oscillatory epoch the scale factor (\ref{aosc})
consists of a linear expansion plus an oscillatory term 
which causes the Hubble parameter (\ref{Hosc}) to become 
negative for brief periods. Because we wish to quantify 
the potential enhancement from these periods of negative 
$H(t)$, it is useful to factor out the behaviour that would 
arise from the linear growth, without the oscillatory term:
\begin{equation}
{\bar a} (t) \; = \;
a_{\rm cr} \Bigl[ 1 + \omega \Delta t \Bigr]
\;\; . \label{alin}
\end{equation}
Then, the Hubble parameter ${\bar H}$ can be expressed 
in terms of the scale factor ${\bar a}$ as follows:
\begin{equation}
{\bar H}(t) \; = \; 
\frac{\omega}{1 + \omega \Delta t} 
\; = \;
\frac{\omega \, a_{\rm cr}}{{\bar a}(t)}
\;\; . \label{Hlin}
\end{equation}
The canonically normalized, Bunch-Davies mode function 
for the linear expansion is \cite{JMPW}:
\begin{equation}
{\bar u}(t,k) \; = \;
\frac1{\sqrt{2 \, \sqrt{c^2 k^2 - \omega^2 a_{\rm cr}^2 } 
\, } \, } \times 
\frac1{{\bar a}(t)} \, \exp \! \Bigg( \!\! 
-i \; \sqrt{\frac{c^2 k^2}{\omega^2 a_{cr}^2} - 1} \,\,
\ln \! \Big[ \, \frac{{\bar a}(t)}{a_{\rm cr}} \,
\Big] \Bigg)
\;\; . \label{ulin}
\end{equation}
Since the product $\, {\bar H}(t) \times {\bar a}(t) \,$ 
is constant, there is no horizon crossing during linear 
expansion. Modes which are sub-horizon at criticality 
($k > \omega a_{\rm cr}$) remain sub-horizon, and 
modes which are super-horizon at criticality ($k < 
\omega a_{\rm cr}$) also remain super-horizon. One 
can see from expression (\ref{ulin}) that sub-horizon 
mode functions oscillate and redshift, with the period 
of oscillation also redshifting. By contrast, 
super-horizon mode functions fall off like:
\begin{equation}
{\bar u}_{\rm super}(t, k) \; \sim \;
\left[ \frac{a_{\rm cr}}{{\bar a}(t)}
\right]^{1 \, \pm \, 
\sqrt{1 - \frac{c^2 k^2}{\omega^2 a_{\rm cr}^2}}}
\;\; . \label{superfall}
\end{equation} 

${\bullet \; \;}$ {\bf Initial Conditions:}
We still have to include the effect of oscillations, 
which can only be done numerically. That defines 
a mode function $\widetilde{u}(t,k)$ which obeys 
equation (\ref{ueqn}) with the full oscillating 
geometry (\ref{aosc}-\ref{Hosc}). We construct 
these mode functions to agree initially with those 
of the fictitious phase of linear expansion:
\begin{equation}
\widetilde{u}(t_{\rm cr},k) \; = \;
{\bar u}(t_{\rm cr},k) 
\qquad , \qquad
\dot{\widetilde{u}}(t_{\rm cr},k) \; = \;
\dot{{\bar u}}(t_{\rm cr},k) 
\;\; . \label{init}
\end{equation}
The actual mode function $u(t,k)$ -- by which we mean 
the coefficient of the annihilation operator in the free 
field expansion -- is neither $\widetilde{u}(t,k)$ nor 
$\widetilde{u}^*(t,k)$, but rather a linear combination 
of the two solutions:
\begin{equation}
u(t,k) \; = \; 
\alpha \; \widetilde{u}(t,k) \, + \,
\beta \; \widetilde{u}^*(t,k)
\;\; . \label{comb}
\end{equation}
We can solve for the combination coefficients in terms 
of the values of $u(t,k)$ and its first derivative at 
criticality:
\footnote{The Wronskian of the two solutions is:
$\widetilde{u} \, \dot{\widetilde{u}}^* - 
\dot{\widetilde{u}} \, \widetilde{u}^* 
= i \, a^{-3}$.}
\begin{eqnarray}
\alpha &\!\! = \!\!& 
-i \, a_{\rm cr}^3 \Bigl[ \,
u(t_{\rm cr},k) \; \dot{\bar u}^*(t_{cr},k)
\, - \,
\dot{u}(t_{\rm cr},k) \; {\bar u}^*(t_{cr},k) 
\, \Bigr] 
\;\; , \label{alpha} \\
\beta &\!\! = \!\!& 
-i \, a_{\rm cr}^3 \Bigl[ \,
\dot{u}(t_{\rm cr},k) \; {\bar u}(t_{cr},k)
\, - \,
u(t_{\rm cr},k) \; \dot{\bar u}(t_{cr},k) 
\, \Bigr] 
\;\; . \label{beta}
\end{eqnarray}

Although we do not know precisely what these values 
are, some reasonable guesses can be made. For example, 
a far super-horizon mode, which experienced first 
horizon crossing $N_{\rm hor}$ e-foldings before 
criticality, should have:
\begin{eqnarray}
u(t_{\rm cr},k) &\!\! \approx \!\!& 
\frac{H_{N_{\rm hor}}}{\sqrt{2 k^3}}
\;\; , \label{superu} \\
\dot{u}(t_{\rm cr},k) &\!\! \approx \!\!& 
- \frac{H^2_{N_{\rm hor}}}{\sqrt{2 k^3}}
\left( \frac{k}{H_{N_{\rm hor}} \; a_{\rm cr}} 
\right)^2 
\left[ \, 1 \, + \, \frac{ik}{H_{N_{\rm hor}} \; 
a_{\rm cr}} \, \right]
\;\; . \label{superudot}
\end{eqnarray}
Recall that the potentially observable modes in the 
cosmic microwave background correspond to $N_{\rm hor} 
\approx 60$, which implies:
\begin{equation}
\frac{k}{H_{N_{\rm hor}} \; a_{\rm cr}} \sim 10^{-26}
\;\; \label{supersmall}
\end{equation}
There is absolutely no point in retaining such small 
numbers. So during the oscillatory phase after 
criticality, the mode function of a cosmologically
observable wave number would be unchanged from 
(\ref{superu}), for all practical purposes. That 
had to be true because, for far super-horizon wave 
numbers, (\ref{ueqn}) simplifies to:
\begin{equation}
\ddot{u}(t,k) \, + \, 3 H(t) \, \dot{u}(t,k) 
\; \approx \; 0
\;\; . \label{supereqn}
\end{equation}
and $\, u(t,k) = constant \,$ remains a solution -- 
independent of $a(t)$ -- for as long as it is valid 
to neglect the last term of (\ref{ueqn}). \\

\begin{figure}
\centerline{\epsfig{file=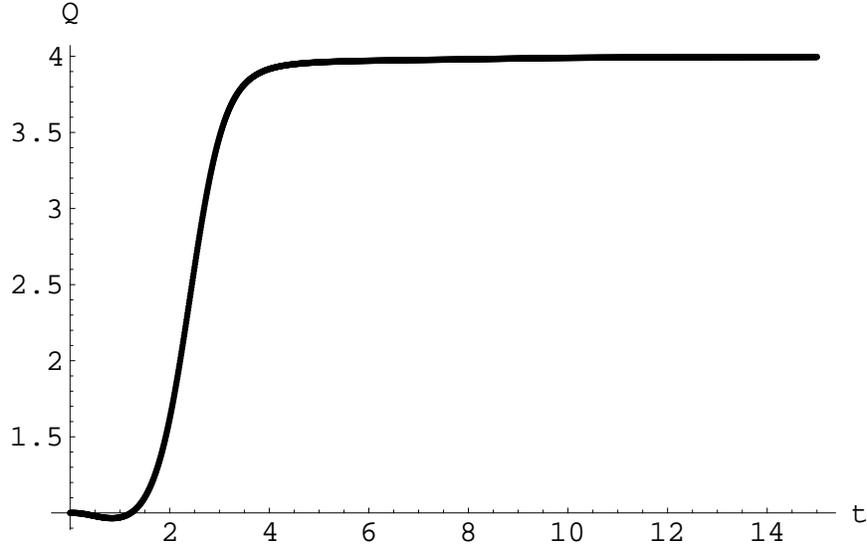,height=2.9in}}
\caption{\footnotesize The enhancement factor $Q$ 
{\it versus} co-moving time (in units of $\omega^{-1}$)
for a
\break \mbox{} \hspace{1.95cm}
super-horizon mode with $\, ck = 0.01 \times 
\omega a_{\rm cr}$.}
\label{grv1}
\end{figure}

${\bullet \; \;}$ {\bf Gravitational Waves Enhancement:} 
Now let us consider the effect of adding the oscillatory 
term to the scale factor. It is intuitively obvious that 
one gets a significant response at resonance; then, the 
natural time scale of the mode function is close to the 
inverse of the oscillatory frequency $\omega$. Whether 
or not this occurs depends upon two things: the wave 
number $k$ and the values of $u(t,k)$ and $\dot{u}(t,k)$ 
at the start of the oscillatory period. The reason the 
initial condition matters is that there are always two,
linearly independent solutions to the mode equation and 
they can have vastly different natural time scales. There 
are three interesting wave number regimes: \\ 

${\ast \;}$ The {\it far super-horizon}, with 
$\, c \, k \ll \omega \, a_{\rm cr}$. \\
From expression (\ref{superfall}) it is evident that, 
without the oscillatory term, super-horizon modes fall 
off with time scales:
\begin{equation}
T_{\pm} \; \simeq \;
\frac{\omega^{-1}}
{1 \pm \sqrt{1 - \frac{c^2 k^2}{\omega^2 a_{\rm cr}^2}} 
\, }
\;\; . \label{superfall2}
\end{equation}
For $\, c k \ll \omega a_{\rm cr} \,$ one of these is 
-- within a factor of two -- close to $\omega$ while 
the other is vastly longer. Numerical analysis shows 
-- see Figure \ref{grv1} -- that the oscillations amplify 
the solution with the shorter time scale by about a 
factor of four. As might be expected, the solution 
with the longer time scale experiences no significant 
amplification. Because the natural initial conditions 
(\ref{superu}-\ref{superudot}) imply the mode enters 
the oscillatory epoch almost entirely in the long time 
scale solution, the effect is that far super-horizon 
modes experience no significant enhancement from the 
oscillation. \\ 

${\ast \;}$ The {\it far sub-horizon}, with 
$\, c \, k \gg \omega \, a_{\rm cr}$. \\
Far sub-horizon modes also receive no substantial 
enhancement, but for a different reason. For far 
sub-horizon modes the natural frequencies of both 
solutions are about $\, c k \, a^{-1}_{\rm cr} \,$
which is much bigger than $\omega$, so neither 
solution experiences much enhancement and it does 
not matter much what the initial condition is. \\

\begin{figure}
\centerline{\epsfig{file=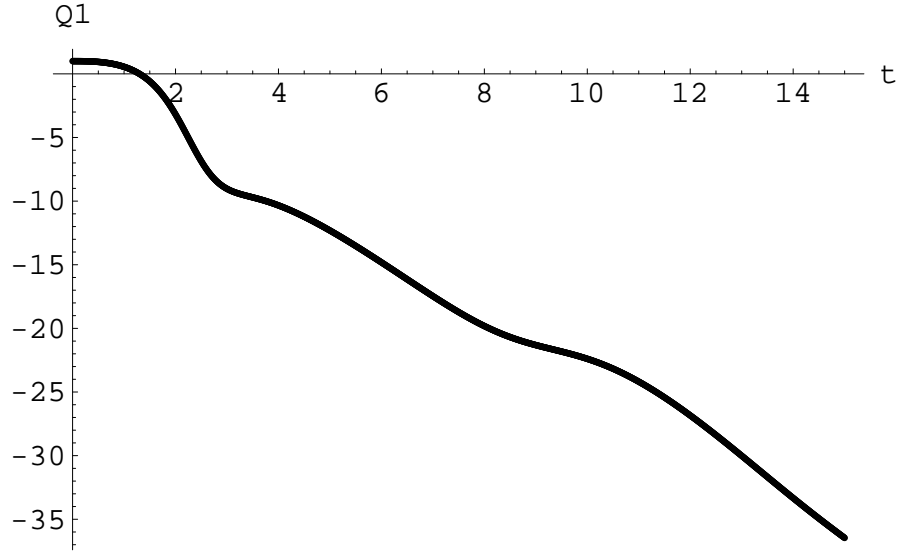,height=2.9in}}
\caption{\footnotesize The enhancement factor $Q1$ 
{\it versus} co-moving time (in units of $\omega^{-1}$)
for a
\break \mbox{} \hspace{1.95cm}
near-horizon mode with $\, ck = 1.1 \times
\omega a_{\rm cr}$.}
\label{grv2}
\end{figure}

\begin{figure}
\centerline{\epsfig{file=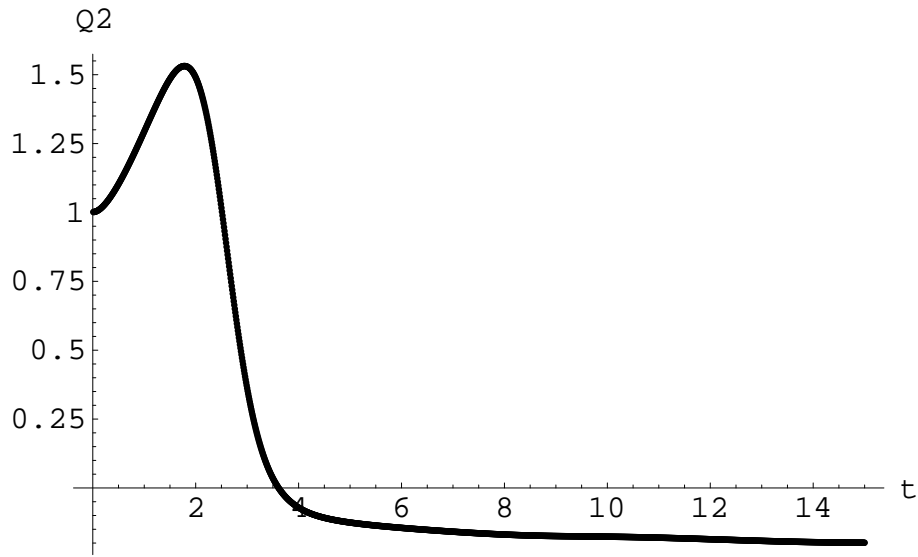,height=2.9in}}
\caption{\footnotesize The enhancement factor $Q2$ 
{\it versus} co-moving time (in units of $\omega^{-1}$)
for a
\break \mbox{} \hspace{1.95cm}
near-horizon mode with $\, ck = 1.1 \times
\omega a_{\rm cr}$.}
\label{grv3}
\end{figure}

${\ast \;}$ The {\it near-horizon}, with 
$\, c \, k \approx \omega \, a_{\rm cr}$. \\
As one might expect, it is the near horizon modes which 
experience the greatest enhancement. Figures \ref{grv2} 
and \ref{grv3} present numerical simulations for the 
case of $\, ck = \frac{11}{10} \omega a_{\rm cr} \,$,
giving the ratios of the actual mode functions -- 
evolved with the oscillatory term -- compared with 
the solution (\ref{ulin}) which starts from the 
same initial condition but is evolved without the 
oscillatory term.
\footnote{The enhancement factor $Q1$ is associated
with the real part: 
$Q1 \equiv {\rm Re} [\widetilde{u}(t, k)] \div 
{\rm Re}[{\bar u}(t, k)]$ 
while $Q2$ with the imaginary part: 
$Q2 \equiv {\rm Im} [\widetilde{u}(t, k)] \div 
{\rm Im}[{\bar u}(t, k)]$.}
In the near horizon regime one expects both solutions 
to be present with about the same amplitude, so a 
reasonable estimate of the total enhancement is by 
adding the two solutions in quadrature and taking 
the ratio with, and without the oscillatory term:
\begin{equation}
Q \; \equiv \;
\frac{\vert \, \widetilde{u}(t, k) \, \vert}
{\vert \, {\bar u}(t, k) \, \vert}
\; \; . \label{Q}
\end{equation}

\begin{figure}
\centerline{\epsfig{file=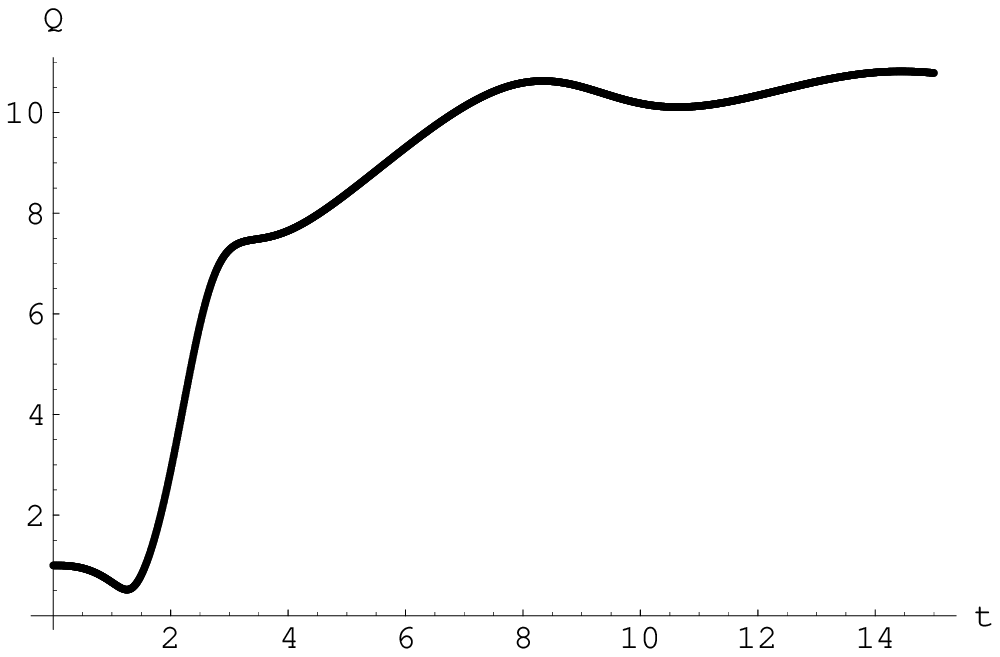,height=2.9in}}
\caption{\footnotesize The enhancement factor $Q$ 
{\it versus} co-moving time (in units of $\omega^{-1}$)
for a
\break \mbox{} \hspace{1.95cm}
near-horizon mode with $\, ck = 1.1 \times
\omega a_{\rm cr}$.}
\label{grv4}
\end{figure}

\begin{figure}
\centerline{\epsfig{file=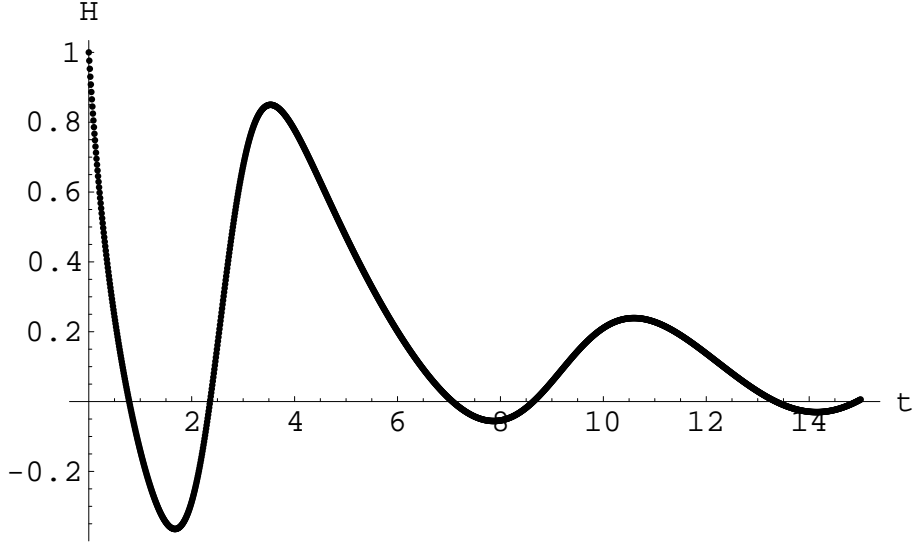,height=2.9in}}
\caption{\footnotesize The Hubble paramemter $H$ 
{\it versus} co-moving time (in units of $\omega$).
The first
\break \mbox{} \hspace{1.9cm}
period of $H < 0$ coincides with the largest growth
of $Q$ in Figure~\ref{grv4} .}
\label{grv5}
\end{figure}
From Figure \ref{grv4} one can see that the enhancement 
factor is about $\, Q \approx 10$. By comparison with 
the Hubble parameter -- see Figure \ref{grv5} -- we see 
that almost all the enhancement derives from the first 
oscillation. This is important because, as we explained, 
the phase of oscillations is likely to be short. The fact 
that almost all the enhancement occurs during the first 
oscillation means that the effect is reliable even if 
only one oscillation occurs.

\begin{figure}
\centerline{\epsfig{file=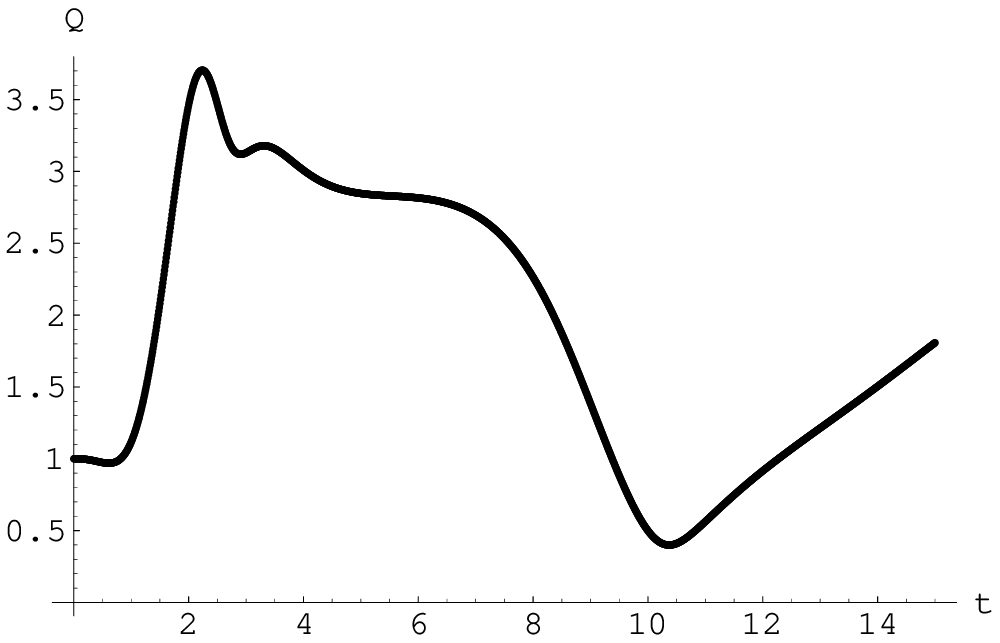,height=2.9in}}
\caption{\footnotesize The enhancement factor $Q$ 
{\it versus} co-moving time (in units of $\omega^{-1}$)
for a
\break \mbox{} \hspace{1.95cm}
near-horizon mode with $\, ck = 2 \times
\omega a_{\rm cr}$.}
\label{grv6}
\end{figure}

\begin{figure}
\centerline{\epsfig{file=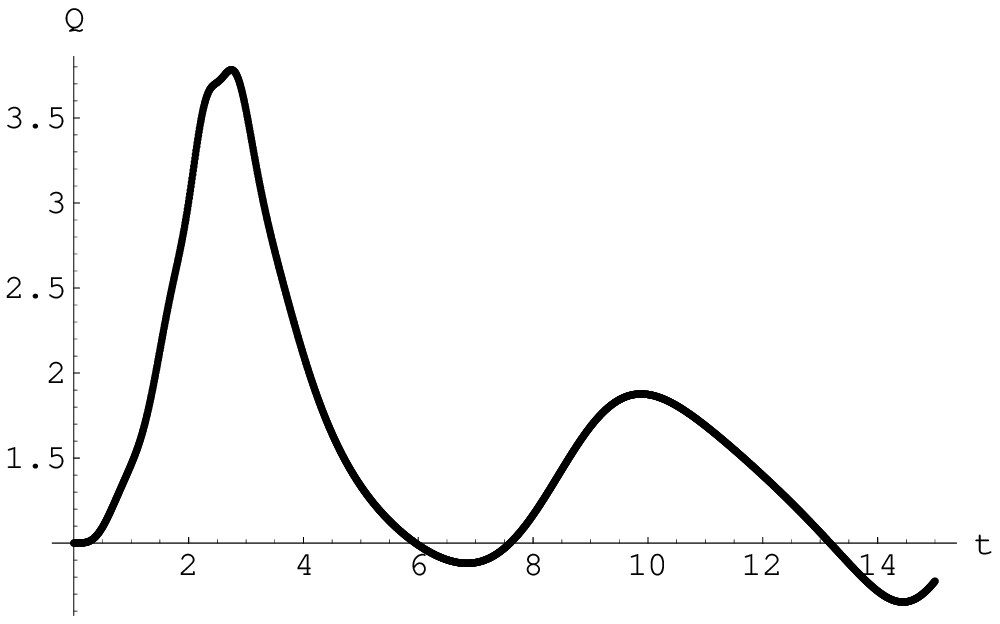,height=2.9in}}
\caption{\footnotesize The enhancement factor $Q$ 
{\it versus} co-moving time (in units of $\omega^{-1}$)
for a
\break \mbox{} \hspace{1.95cm}
near-horizon mode with $\, ck = 5 \times
\omega a_{\rm cr}$.}
\label{grv7}
\end{figure}

Figures \ref{grv6} and \ref{grv7} give the total enhancement 
factor for the cases of $\, ck = 2 \omega a_{\rm cr} \,$ 
and $\, ck = 5 \omega a_{\rm cr} \,$, respectively. Note 
that the enhancement falls off rapidly as one moves away 
from resonance. Note also that the enhancement factor 
oscillates. 

\begin{figure}
\centerline{\epsfig{file=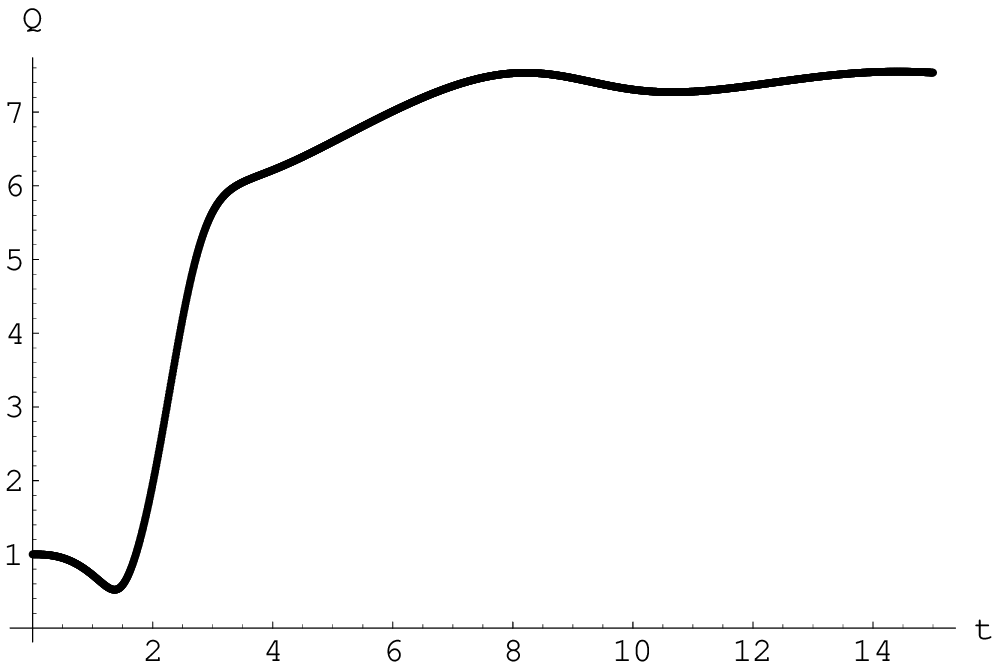,height=2.9in}}
\caption{\footnotesize The enhancement factor $Q$ 
{\it versus} co-moving time (in units of $\omega^{-1}$)
for a
\break \mbox{} \hspace{1.95cm}
near-horizon mode with $\, ck = 0.9 \times
\omega a_{\rm cr}$.}
\label{grv8}
\end{figure}

\begin{figure}
\centerline{\epsfig{file=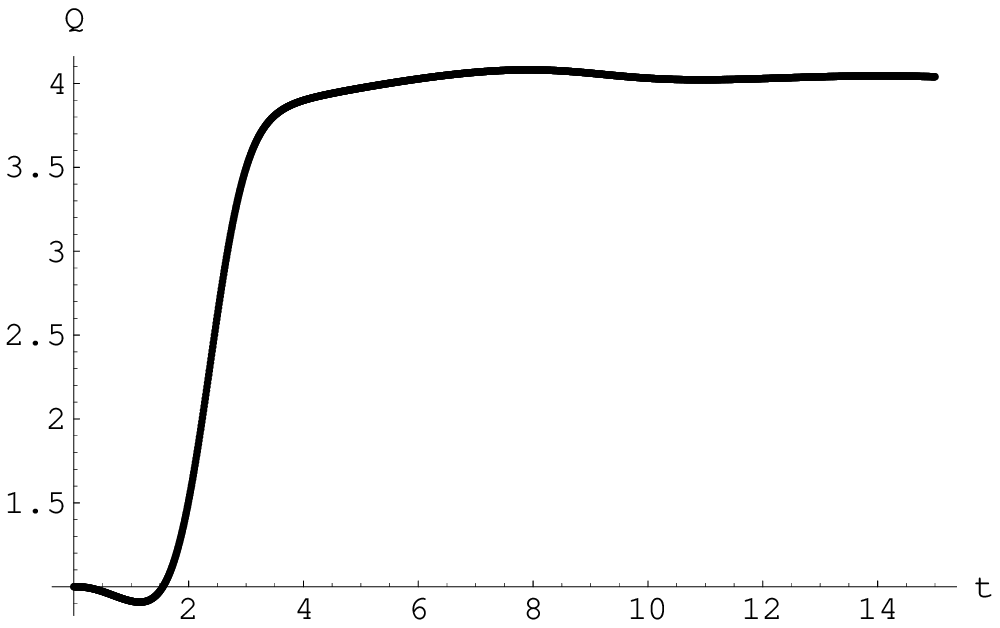,height=2.9in}}
\caption{\footnotesize The enhancement factor $Q$ 
{\it versus} co-moving time (in units of $\omega^{-1}$)
for a
\break \mbox{} \hspace{1.95cm}
near-horizon mode with $\, ck = 0.5 \times
\omega a_{\rm cr}$.}
\label{grv9}
\end{figure}

\begin{figure}
\centerline{\epsfig{file=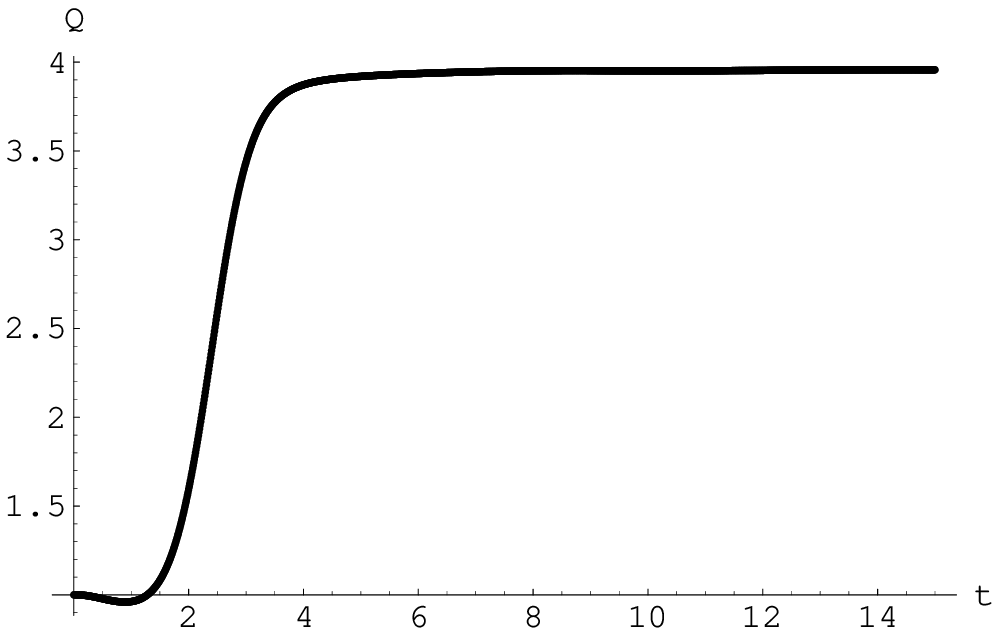,height=2.9in}}
\caption{\footnotesize The enhancement factor $Q$ 
{\it versus} co-moving time (in units of $\omega^{-1}$)
for a
\break \mbox{} \hspace{1.95cm}
near-horizon mode with $\, ck = 0.2 \times
\omega a_{\rm cr}$.}
\label{grv10}
\end{figure}

Modes which are slightly super-horizon are quite similar 
to those which are slightly sub-horizon. Figure \ref{grv8} 
gives the total enhancement factor for the case of 
$\, ck = \frac{9}{10} \omega a_{\rm cr}$. However, 
decreasing the wave number much more rapidly reaches 
the factor of four enhancement which is concentrated 
on the solution that is not likely to be present. 
Figures \ref{grv9} and \ref{grv10} give the behaviours 
for $\, ck = \frac12 \omega a_{\rm cr} \,$ and 
$\, ck = \frac15 \omega a_{\rm cr} \,$, respectively. \\

${\bullet \; \;}$ {\bf Enhanced Waves Energy Density 
and Frequency:} 
It remains to estimate the current energy density and 
frequency of gravitons which are produced during the 
epoch of oscillations. Suppose we regard the enhancement 
factor as $\, Q = 10 \,$ for modes within the range 
$\, \frac23 \omega a_{\rm cr} < ck < 
\frac32 \omega a_{\rm cr} \,$, and zero outside this 
band. This is superimposed on the mode functions 
(\ref{ulin}) of linear expansion. A reasonable estimate 
for the extra physical energy -- above the 0-point value 
of $\, \frac12 \hbar c k \times a^{-1}(t)$ -- in a single 
wave vector $\bf{k}$ within the band of enhancement is:
\begin{equation}
E(t_{\rm cr},k) \; \sim \;
\frac{1}{a_{\rm cr}} \,\hbar c^2 k^2 \, 
\Big\vert Q \; {\bar u}(t_{\rm cr}, k) \Big\vert^2 
\; = \;
\frac{Q^2 \, \hbar c^2 k^2}
{2 \omega a^2_{\rm cr} 
\sqrt{\vert 1 - (\frac{c k}{\omega a_{\rm cr}})^2 \vert} \,} 
\;\; . \label{E1}
\end{equation}

\newpage

\begin{figure}
\centerline{\epsfig{file=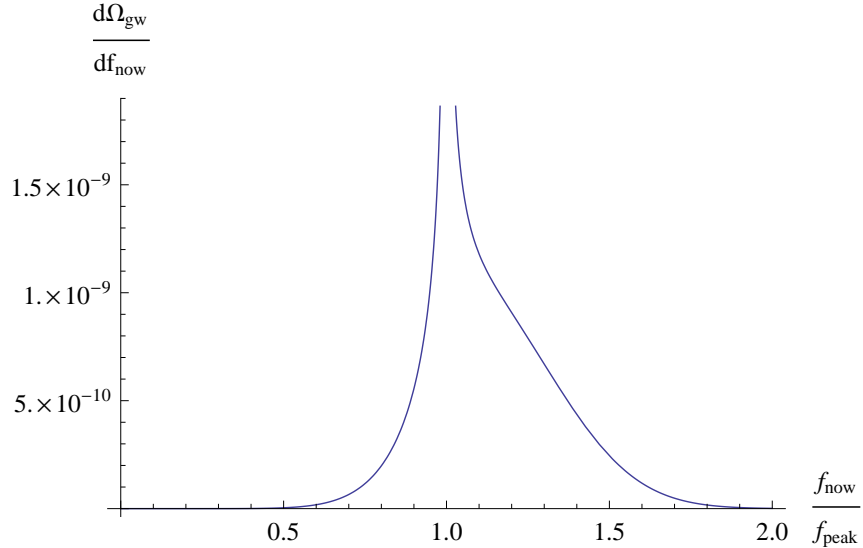,height=2.9in}}
\caption{\footnotesize Fraction of the current energy 
density in gravity waves from our signal per
\break \mbox{} \hspace{1.95cm}
frequency.}
\label{grv11}
\end{figure}

The associated energy density comes from integrating 
the extra physical energy (\ref{E1}) over all the modes 
whose wave numbers are within the band of enhancement:
\begin{eqnarray}
\rho_{\rm gw}(t_{\rm cr}) &\!\! \sim \!\!&
\frac1{a_{\rm cr}^3} \int \frac{d^3k}{(2\pi)^3} \;
\theta \Big( \, \frac32 \, \omega a_{\rm cr} - ck \Big) \;
\theta \Big( ck - \frac23 \, \omega a_{\rm cr} \Big) \;
E(t_{\rm cr}, k) 
\label{rhocr2} \\
&\!\! \sim \!\!&
\frac{\, Q^2 \, \hbar \omega^4}{4 \pi^2 c^3} 
\;\; . \label{rhocr1} 
\end{eqnarray}
Perhaps of more relevance for gravity waves detectors is
the amount of energy density per unit frequency since such 
detectors are only sensitive in certain frequency bands. 
We can estimate the energy density per angular frequency 
at the critical time by simply not performing the radial 
integration over $k$ in expression (\ref{rhocr1}):
\begin{equation}
\frac{ d\rho_{\rm gw}(t_{\rm cr})}{d ck} \; \sim \;
\frac{100}{4 \pi^2} \,
\frac{\hbar \omega^3}{c^3 a_{\rm cr}} \,
\frac{ (\frac{Q}{10})^2 \,(\frac{ck}{\omega a_{\rm cr}})^4 }
{ \sqrt{\vert 1 - (\frac{ck}{\omega a_{\rm cr}})^2 \vert} } 
\;\; . \label{rhoperfreq} 
\end{equation}
To convert this to the current epoch we note that these 
gravitons are sub-horizon after the phase of oscillations, 
so their energy density redshifts like radiation:
\begin{equation}
\rho_{\rm gw}(t_{\rm now}) \; = \;
\Bigl(\frac{a_{\rm cr}}{a_{\rm now}} \Bigr)^4 
\rho_{\rm gw}(t_{\rm cr}) 
\;\; . \label{rhoredshift}
\end{equation}
We also take note of the relations between the ordinary 
(not angular) current frequency and peak frequency:
\begin{equation}
f_{\rm now} \, = \,
\frac{c k}{2\pi a_{\rm now}} \qquad , \qquad 
f_{\rm peak} \, = \,
\frac{\omega a_{\rm cr}}{2\pi a_{\rm now}} 
\;\; , \label{fnowpeak}
\end{equation}
and we divide out a factor of the current critical 
density to obtain the fraction in gravity waves:
\begin{equation}
\Omega_{\rm gw} \, \equiv \,
\rho_{\rm gw}(t_{\rm now}) \div 
\Bigl(\frac{3 c^2 H^2_{\rm now}}{8 \pi G} \Bigr) 
\;\; . \label{Omegagw}
\end{equation}
The final result is:
\begin{equation}
\frac{d \Omega_{\rm gw}}{d (\frac{f_{\rm now}}{f_{\rm peak}})} 
\; \sim \;
\Bigl( \frac{200}{3\pi} \Bigr) 
\Bigl( \frac{\hbar G \omega^2}{c^5} \Bigr)
\Bigl( \frac{\omega}{H_{\rm now}} \Bigr)^2 
\Bigl(\frac{a_{\rm cr}}{a_{\rm now}}\Bigr)^4 \times 
\frac{(\frac{Q}{10})^2 \, (\frac{f_{\rm now}}{f_{\rm peak}})^4}
{\sqrt{\vert 1 - (\frac{f_{\rm now}}{f_{\rm peak}})^2 \vert}}
\;\; . \label{graph11a}
\end{equation}
Using the restrictions (\ref{Nnowval}, \ref{omegabound}) 
gives:
\begin{equation}
\frac{d \Omega_{\rm gw}}{d (\frac{f_{\rm now}}{f_{\rm peak}})} 
\; \sim \;
\Bigl( 4 \times 10^{-10} \Bigr) \times 
\frac{(\frac{Q}{10})^2 \, (\frac{f_{\rm now}}{f_{\rm peak}})^4}
{\sqrt{\vert 1 - (\frac{f_{\rm now}}{f_{\rm peak}})^2 \vert}}
\;\; . \label{graph11b}
\end{equation}
A rough model for the $Q$-factor supported by our 
numerical analysis is:
\begin{equation}
\frac{Q}{10} \; \sim \;
\exp \! \left[ [-4 \left( 
\frac{f_{\rm now}}{f_{\rm peak}} - 1 \right)^2 \right] 
\;\; . \label{Qfor graph11}
\end{equation}
and it is for this choice of the $Q$-factor that Figure 
\ref{grv11} gives a plot of $\, [ d\Omega_{\rm gw} / 
d(\frac{f_{\rm now}}{f_{\rm peak}}) ] \,$ as a function 
of $\, [ f_{\rm now} / f_{\rm peak} ]$.

From Figure \ref{grv11} it is evident that the signal is 
highly peaked at the frequency $f_{\rm peak} \sim 10^9 H\!z$, 
and is negligible at significantly different frequencies. 
It would be challenging to detect gravitational radiation 
at such high frequencies but detectors in that range have 
been proposed \cite{MHz}. As noted in the text, the phase 
of oscillations does not affect modes which experienced 
first horizon crossing more than a few e-foldings before 
the end of inflation. The wavelength of our effect is 
$\, \lambda_{\rm peak} = c f^{-1}_{\rm peak} \gtwid 0.3 m$, 
whereas the smallest scale feature which is currently 
observed in the cosmic microwave radiation is about 
$10^{22} m$ \cite{ACBAR}! Our model does not change either 
how matter couples to gravity or the propagation of 
linearized gravitons, so it has no effect on the spin-down 
rate of the binary pulsars. The gravity waves we predict 
will certainly distort how pulsar light propagates, but
the short wavelength again seems to preclude a detectable 
effect. LIGO is not sensitive above frequencies of 
$\, 7000 H\!z$, which is far too low. The situation 
is even worse with LISA's high frequency cutoff of 
$\, 0.1 H\!z$.

\vspace{1cm}

\centerline{\bf Acknowledgements}
We are grateful to B. A. Bassett and S. M. Sibiryakov
for encouraging us to make this study, and to M. Szabolcs
for informing us about the possibility of very high 
frequency gravity wave detectors. 
This work was partially supported by the European Union 
grant FP-7-REGPOT-2008-1-CreteHEPCosmo-228644, 
by NSF grants PHY-0653085 and PHY-0855021, 
and by the Institute for Fundamental Theory at the 
University of Florida.

\end{document}